\documentclass[iop]{emulateapj}
\usepackage{apjfonts}
\usepackage{graphicx}
\usepackage{color}

\usepackage{epstopdf}
\usepackage{amsmath}

\def\ltsima{$\; \buildrel < \over \sim \;$}
\def\simlt{\lower.5ex\hbox{\ltsima}}
\def\gtsima{$\; \buildrel > \over \sim \;$}
\def\simgt{\lower.5ex\hbox{\gtsima}}

\newcommand\lsim{\mathrel{\rlap{\lower4pt\hbox{\hskip1pt$\sim$}}
\raise1pt\hbox{$<$}}}
\newcommand\gsim{\mathrel{\rlap{\lower4pt\hbox{\hskip1pt$\sim$}}
\raise1pt\hbox{$>$}}}

\allowdisplaybreaks
\usepackage{color}

\shorttitle{It's Raining Black Holes}
\shortauthors{Naoz \& Haiman }
\begin{document}

\title{ The Enhanced Population of Extreme Mass-Ratio Inspirals in the LISA Band from Supermassive Black Hole Binaries }
%Extreme-Mass-Ratio Inspirals as a Possible Noise Source for LISA

\author{  Smadar Naoz$^{1,2}$, Zolt\'an Haiman$^{3,4}$}

\altaffiltext{1}{Department of Physics and Astronomy, University of California, Los Angeles, CA 90095, USA}
\altaffiltext{2}{Mani L. Bhaumik Institute for Theoretical Physics, Department of Physics and Astronomy, UCLA, Los Angeles, CA 90095, USA}
\altaffiltext{3}{Department of Astronomy, Mail Code 5246, Columbia University, New York, NY 10027, USA}
\altaffiltext{4}{Department of Physics, Mail Code 5255, Columbia University, New York, NY 10027, USA}

\begin{abstract}
  Extreme mass ratio inspirals (EMRIs) take place when a stellar-mass black hole (BH) merges with a supermassive black hole (SMBH). The gravitational wave emission from such an event is expected to be detectable by the future  Laser Interferometer Space Antenna (LISA) and other mHz detectors.  It was recently suggested that the EMRI rate in SMBH binary systems is orders of magnitude higher than the EMRI rate around a single SMBH with the same total mass.  Here we show that this high rate can produce thousands of SMBH-BH sources at a redshift of unity. We predict that LISA may detect a few hundred of these EMRIs with signal-to-noise ratio above SNR$\geq$8 within a four-year mission lifetime.  The remaining sub-threshold sources will contribute to a large confusion noise, which is approximately an order of magnitude above LISA's sensitivity level. Finally, we suggest that the individually detectable systems, as well as the background noise from the sub-threshold EMRIs, can be used to constrain the SMBH binary fraction in the low-redshift Universe.   
\end{abstract}
\maketitle

\section{Introduction}

The gravitational-wave (GW) emission from a merger of a stellar-mass black hole (BH) and a supermassive black hole (SMBH) is one of the science drivers of the Laser Interferometer Space Antenna (LISA), as well as other planned mHz detectors, such as TianQin \citep[e.g.,][]{Amaro-Seoane+17,Robson+19,Baker+19,Mei+20}.  Thus, the formation channel and rate of these extreme-mass-ratio inspirals (EMRIs) are of prime importance for these future missions.

One of the main EMRI formation channels is based on the weak gravitational interactions between neighboring stellar-mass BHs surrounding SMBHs at the centers of galaxies. These weak two-body interactions are referred to as ``two-body relaxation'' and describe the so-called ``loss-cone'' dynamics in dense nuclear star clusters \citep[e.g.,][]{Merritt10,Binney+TremaineBook}. In  this channel, the encounters in the dense cluster surrounding the SMBH can lead to extreme eccentricity for a stellar mass BH, driving it toward the SMBH at the center of the cluster \citep[e.g.,][]{Hopman+05,Aharon+16,Amaro18,Sari+19}. This scenario yields an EMRI rate of a few tens of EMRIs per Gyr per galaxy \citep[e.g.,][]{Hopman+05,Aharon+16}\footnote{Note that collisions between stellar mass BHs and the surrounding stars can lead to a build-up of the stellar BH's mass to reach the intermediate-mass BH regime \citep{Rose+22}.}. 

Recently, \citet[][hereafter N22]{Naoz+22} suggested that EMRI formation in an SMBH binary configuration is an extremely efficient process. SMBH binaries are a natural consequence of the hierarchical formation of galaxies \citep[e.g.,][]{DiMatteo+05,Hopkins+06,Robertson+06,Callegari+09,Li+20Pair}. This concept is supported by AGEN pairs and observations of SMBHs on wide orbits \citep[e.g.,][]{Komossa+08,Bianchi+08,Comerford+09bin,Green+10,Smith+10,Caproni+17,Stemo+20}. These observations imply that those systems may evolve to become tighter (reaching separations on the order of a parsec) binaries. A handful of promising individual candidates on sub-parsec orbits have been reported in the literature \citep[e.g.,][]{Sillanpaa+88,Rodriguez+06,Bansal+17,Ren+21,ONeill+22}, and samples of several dozen quasars have been identified as sub-parsec binary candidates based on periodicities of their light-curves in large time-domain surveys~\citep{Graham+2015,Charisi+2016,Liu+2019,Chen+2020}. Our own Galactic center environment, too, is consistent with hosting a possible less massive intermediate-mass black hole companion to Sgr A* \citep[e.g.,][]{Hansen+03,Maillard+04,Grkan+05,Gualandris+09,Generozov+20,Naoz+20,Gravity+20,Fragione+20,Zheng+20,Rose+22,Zhang+23,Gravity+23IMBH}.

In the N22 scenario, the SMBH undergoing a high EMRI rate is the less massive of the SMBH pair. As was shown in their study, the eccentric Kozai Lidov \citep[EKL;][]{Kozai,Lidov,Naoz16} mechanism, combined with two body relaxation, is more efficient around the lighter, secondary SMBH than around the more massive primary one. This behavior is a result of general relativistic (GR) precession, which strongly suppresses eccentricity excitations around the primary SMBH, but less so around the secondary \citep[e.g.,][]{Ford00,Naoz+13GR,Naoz+14,Li+15}. Eccentricity excitations still take place around the secondary, however, less efficiently and typically via scattering events \citep[e.g.,][]{Iva+10,Chen+09,Chen+11,Bode+14}.  Specifically, N22 demonstrated that the combined channels of EKL and two-body relaxation are vital to producing an efficient rate. They showed that in this mechanism, all available stellar-mass BHs around the lighter SMBH are driven to high eccentricity yielding a GW source in this process.

This channel has similar implications on the rate of tidal disruption events (TDEs). Particularly, it was previously suggested that SMBH binaries can result in a fast, temporary (a few Myrs) increase of the TDE rate \citep[e.g.,][]{Iva+10,Chen+09,Chen+11,Wegg+11,Bode+14,Li+15,Fragione+18,li_direct_2019}. However, most of the investigation in the literature focused on disrupting stars on the primary (the central, more massive SMBH) due to perturbations from the secondary (the less massive one). In this scenario, EKL eccentricity excitations are suppressed due to general relativity precession \citep[e.g.,][]{Naoz+13GR} and the majority of TDEs take place due to chaotic orbital crossings \citep[e.g.,][]{Chen+09,Chen+11,Wegg+11}. Focusing on TDEs on the secondary (less massive one) allows for efficient EKL excitations to produce TDEs \citep{Li+15}, which can be used to detect the lighter SMBH in the binary pair \citep{Mockler+23}. Including two-body relaxation to the EKL yields an efficient process similar to the EMRI case. Specifically, it was shown by \citet{Melchor+23} that it increases the number of stars that undergo TDEs, over a longer period of time.  As a result, the TDE rate is consistent with the observed rate in post-star burst (E+A) galaxies \citep[see Figure 9, left panel in][]{Melchor+23}. This scenario also relies on having an unequal-mass SMBH binary configuration.

\begin{figure}
  \begin{center} %\vspace{-1.2cm} %\hspace{0.2cm} 
\includegraphics[width=\linewidth]{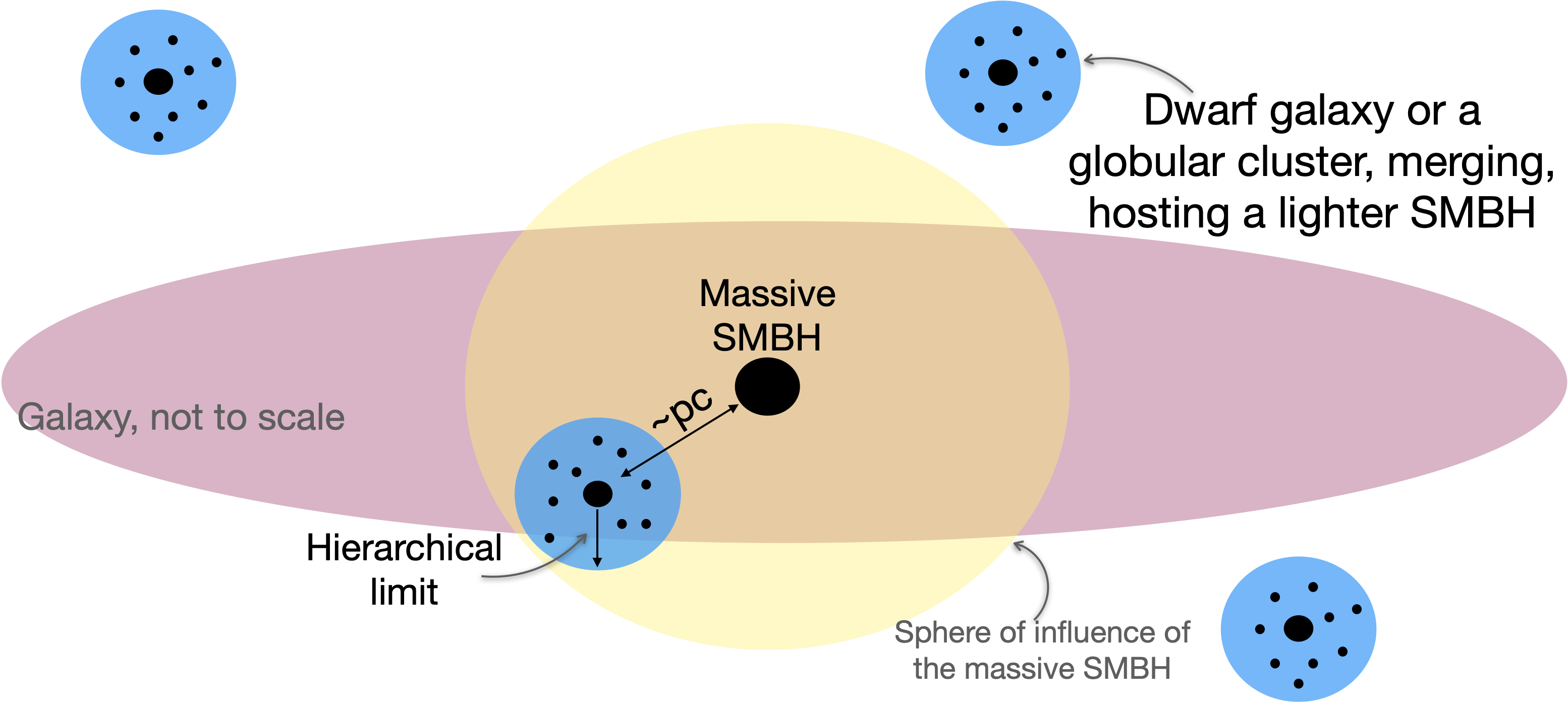}
  \end{center} %\vspace{-0.4cm} %\hspace{0.2cm}
  \caption{  \upshape {\bf{ An illustration of the system.}} The hierarchical nature of galaxy formation suggests that a galaxy will undergo multiple mergers with smaller structures, such as dwarf galaxies or global clusters. If these smaller objects host SMBHs (or IMBHs) at their centers, the central galaxy's SMBH can have several lower-mass SMBH (or IMBH) companions.}
  \label{fig:cartoon}
\vspace{0.4cm}
\end{figure}

Interestingly, the hierarchical nature of galaxy formation suggests that a massive SMBH may have a high probability of hosting lighter companions. Since mergers with dwarfs or smaller galaxies are more probable for a given galaxy \citep[e.g.,][]{Volonteri+03,Rodriguez-Gomez+15,Pillepich+17,Paudel+18,Micic+23}, an SMBH in the center of a galaxy may have multiple smaller SMBH companions throughout the lifetime of the galaxy \citep[either one at a time or a few at a time, e.g.,][]{Madau+01,Volonteri+03cores,Rashkov+14,Ricarte+16}.  See Figure \ref{fig:cartoon} for an illustration of this arrangement. This configuration was adopted in the aforementioned study by N22, who estimated the EMRI rate to be on the order of $\sim 10^3$ to $\sim 10^5$ EMRIs per primary SMBH (for SMBH masses from $\sim 10^6$ to $\sim 10^8\,{\rm M_\odot}$).
%It had been suggested that since lower mass dwarf galaxies take longer to merge with a large galaxy, the dwarfs' low-mass SMBHs will tend to cluster in the bulges of large galaxies \citep[e.g.,][]{Madau+01,Volonteri+03}. However, other arguments and high-resolution simulations showed that in some cases, the dwarfs' SMBHs may sink all the way to the central parts of the galaxy, holding to a relic of stars within their own sphere of influence   \citep[e.g.,][]{Rashkov+14,Lacroix+18,Boldrini+20,Greene+20,Chu+23}. In such a case, EMRI rate should be significantly enhanced.
It had been suggested that lower-mass dwarf galaxies take longer to merge with a large galaxy, and suffer tidal stripping which reduces their mass and increases the dynamical friction infall timescale.   If so, these dwarfs' low-mass SMBHs will wander away from the large galaxies' center \citep[e.g.,][]{Madau+01,Volonteri+03}. However, high-resolution simulations, including gaseous dissipation, showed that if the low-mass companion galaxy is gas-rich, its SMBHs may sink all the way to the central parts of the galaxy, also holding on to a relic of stars within their own sphere of influence  \citep[e.g.,][]{Kazantzidis+2005,Callegari+09,Rashkov+14,Lacroix+18,Boldrini+20,Greene+20,Chu+23}. In these cases, the EMRI rate should be significantly enhanced.

Given this high rate, in this {\it Letter}, we ask: what are the consequences of a high event rate for future GW detectors such as LISA?  In particular, we obtain the boost in the EMRI rate accounting for binaries with different masses, and estimate the possible confusion noise for LISA at the predicted high event rate. While LISA can resolve individual EMRIs, the aforementioned high rate suggests that the unresolved systems can accumulate incoherently to an unresolved confusion noise or a stochastic background. The possible stochastic background from EMRIs was estimated, for the single SMBH case, to produce a low stochastic background, and in some optimistic cases, perhaps a more significant effect is expected \citep[e.g.,][]{Barack+04,Bonetti+20,Pozzoli+23}.  We show and quantify below that the unresolved stochastic background expected from EMRIs in SMBH binaries may be significantly above LISA's sensitivity.

This paper is organized as follows: we begin by using the SMBH mass function in order to estimate the cosmological EMRI rate (\S\ref{sec:rate}). The noise level within LISA is then estimated in \S\ref{sec:LISA}. Lastly, we discuss the astrophysical implications in \S\ref{sec:diss} and offer our conclusions in \S\ref{sec:con}.

\section{Cosmological Rate Estimation}\label{sec:rate}

N22 estimated the EMRI rate per SMBH as a function of the (secondary) SMBH mass $M_\bullet$.
Here we present a fit to their result, $\gamma(M_\bullet)$, as
\begin{equation}\label{eq:gamma}
    \gamma(M_\bullet)\sim 925.4~{\rm Gyr}^{-1} \left(\frac{M_\bullet}{10^6~{\rm M}_\odot}\right)^{0.76}\times f_{\rm LISA}\times f_{\rm EKL} \times \eta_{\rm bin}\ ,
\end{equation}
where the value is based on the $M-\sigma$ relation and a fraction of stellar black holes, of $3.2\times 10^{-3}$ \citep[e.g.,][]{Kroupa01}, which is equal to the total mass of stellar-mass black holes divided by the total mass in stars, within the sphere of influence of an SMBH.  The total mass enclosed within the sphere of influence of an SMBH is twice the SMBH mass \citep[e.g.,][]{Tremaine+02}. In the scenario presented here, each falling SMBH carries its nuclear star cluster with it.  The radius of the secondary's nuclear star cluster shrinks, as the primary's gravitational potential begins to overtake the secondary's. This radius, which is often referred to as the ``hierarchical limit,'' roughly describes the region where the EKL equation of motions can be applied \citep[e.g.,][]{LN11,Naoz16,Zhang+23}. In our case, this is on the order of parsec, which is of the order of the sphere of influence. The number of enclosed stellar mass black holes can be approximated by adopting the $M-\sigma$ relation for these smaller SMBHs' nuclear star clusters. Adopting the value $3.2\times 10^{-3}$ for a fraction of stellar black holes means that we neglect the effects of mass segregation close to the SMBH. Mass segregation may increase the number of stellar BHs, \citep[e.g.,][]{Hopman+06seg}, thus yielding an even larger boost to the EMRI rate than the one described below.

We also introduce the following parameters:
$f_{\rm LISA}$ is the fraction of stellar-mass BHs seen to merge with SMBHs in the LISA band as EMRIs, rather than as plunges. N22 estimated this fraction to be between $40\%-100\%$ (see their Appendix~A).
$f_{\rm EKL}$ is the fraction of stellar-mass BHs in the SMBHs' spheres of influence that are driven to merge with the SMBH when both EKL and two-body relaxation operate. N22 estimated $f_{\rm EKL}\sim 0.5-1$.
Finally, $\eta_{\rm bin}$ is the fraction of SMBHs that acquire a more massive SMBH companion. While this fraction is at most unity, below, we allow $\eta_{\rm bin}$ to be larger than unity due to uncertainties in the SMBH mass function, especially at the low-mass end where it is extrapolated through population modeling, as well as additional uncertainties which are degenerate with this number and may therefore be captured by this single parameter ({{see Section \ref{sec:diss}}}).

Equation (\ref{eq:gamma}) is derived from Figure~5 in N22, which represents a scaling from their fiducial system. It adopts a fixed EKL timescale and rescales the binary separation accordingly. \citet{Mockler+23} and \citet{Melchor+23} have recently showed that keeping the SMBH binary separation constant yields consistent results.
In fact, \citet{Melchor+23}, considering TDEs via the combined EKL and two body relaxation mechanisms, showed that the number of stars (proportional to the primary mass) is the main factor determining the TDE rate. The number of available stars in the hierarchical limit is directly related to the SMBH binary eccentricity, with the lowest eccentricity yielding a higher number of particles and, thus, a higher rate. Since Eq.~(\ref{eq:gamma}) is derived from N22 who adopted $e=0.7$ for their fiducial system, it represents an EMRI rate that must be lower compared to yet smaller SMBH binary eccentricity values.  In other words, this implies that the lower limit of $f_{\rm EKL}$, may be higher than estimated in N22. 
Furthermore, \citet{Melchor+23} considered two mass ratios, $M_2/M_1=10$ and 100, demonstrating that they have little effect on the TDE rate, as long as the disruption is taking place on the less massive SMBH.

Furthermore, N22 and \citet{Melchor+23} demonstrated that the SMBH companion drives the stellar-mass BHs (or stars) toward the primary via this channel on timescales that can be as short as $10^5$~yrs, up to $10^8$~yrs.  The hardening of an SMBH binary may take between millions to billions of years  \citep[e.g.,][]{Begelman+80,Milosavljevic+01,Yu02,Sesana+06,Haiman+2009,Kelley+17}. Note that the hardening process may assist EMRI formation and still achieve a high rate \citep{Mazzolari+22}.

As highlighted above, there are many uncertainties involved in this process. For example, the SMBH binary mass ratio, separation and eccentricity, the hardening rate, and the probability of the SMBH having a companion(s) in the first place. For simplicity, here we fold all of these uncertainties into the parameter $\eta_{\rm bin}$. We offer possible constraints on this parameter in the discussion below.

\begin{figure}
  \begin{center} %\vspace{-1.2cm} %\hspace{0.2cm} 
    \includegraphics[width=\linewidth]{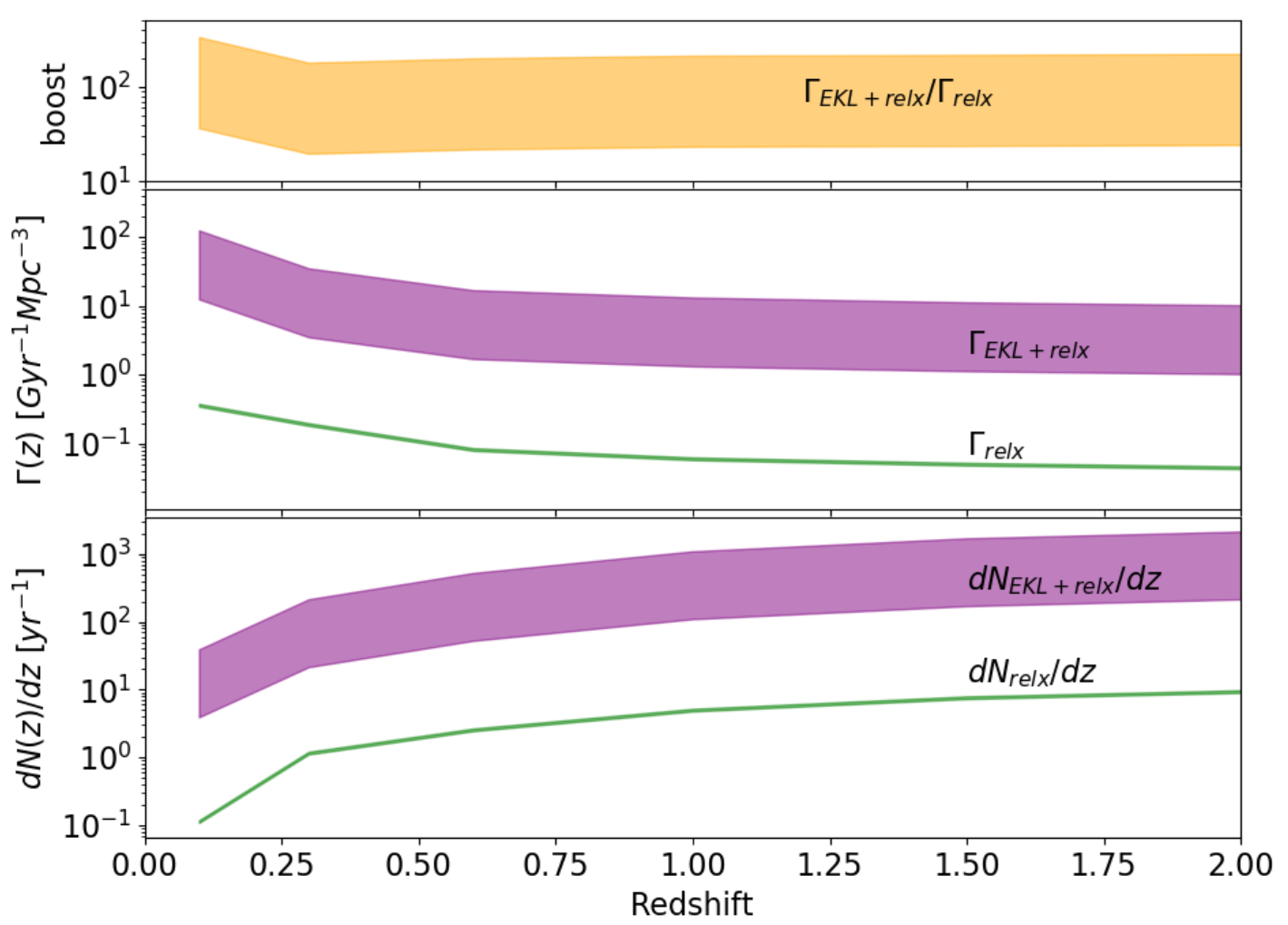}
  \end{center} %\vspace{-0.4cm} %\hspace{0.2cm}
  \caption{  \upshape {\bf Bottom panel:} The expected number of EMRIs per unit redshift per year. We consider the SMBH binary scenario (wide purple band), where the upper limit assumes $\eta_{\rm bin}=1$, and $f_{\rm LISA}=f_{\rm EKL}=1$ and the lower limit assumes $\eta_{\rm bin}=0.5$, $f_{\rm LISA}=0.4$ and $f_{\rm EKL}=0.5$
    \citep[the latter two choices motivated in][]{Naoz+22}. The thin green line shows the corresponding rate for single SMBHs via the two-body relaxation channel. {\bf Middle panel:} the volumetric rate of EMRIs as a function of redshift (see Eq.~\ref{eq:volRate}). The bottom panel is obtained by integrating this rate over volume and allowing one year of LISA operation (Eq.~(\ref{eq:number})). {\bf Top panel:} the enhancement of the EMRI rate in the binary scenario relative to the single-SMBH case.  The numbers in this figure do not take into account LISA's sensitivity (see \S\ref{sec:LISA}). } \label{fig:numbers} %\vspace{-0.4cm}
\end{figure}

\citet{Merloni+08} estimated the SMBH mass function by solving the continuity equation for its evolution, and using the locally determined mass function as a boundary condition \citep[see also][]{Sicilia+22}. Similar estimates were found by \citet{Aversa+15}, who calibrated their results to so-called abundance matching.  In the Appendix, we show the fit we used for the SMBH mass function.  Note that this mass function assumes that the mass of each nucleus is dominated by the more massive (central) SMBH; however, in our scenario, each nucleus may hold more than one SMBH; thus, this mass function undercounts small SMBHs. Given the overall uncertainty in the mass function \citep[e.g.,][]{Merloni+08,Aversa+15,Sicilia+22}, we do not expect this multiplicity to alter our conclusions significantly.  Given the number density $d^2 N(z)/d\log(M_\bullet)dV$ (i.e., the number of SMBHs per comoving volume and mass, as a function of redshift $z$ and mass $M_\bullet$), we can estimate the EMRI rate density at each redshift as
\begin{equation}\label{eq:volRate}
    \Gamma(z) = \int  \frac{d^2N}{dV d\ln M} \gamma(M)d\ln M \ , %\eta_{\rm bin}
\end{equation}
Thus, the number of EMRIs per unit $z$ expected %to be detected by LISA over its 
{{over LISA's}} lifetime $T_{\rm obs}$ is:
\begin{equation}\label{eq:number}
    \frac{dN}{dz}(z) =T_{\rm obs} \times \int \frac{\Gamma(z)}{1+z} \frac{dV}{dz}   \ ,
\end{equation}
where $dV/dz$ is the comoving volume per unit redshift, and the factor $(1+z)^{-1}$ accounts for cosmological time dilation.

In the bottom panel of Figure~\ref{fig:numbers}, we show an example of the number of EMRIs per unit redshift, expected after $T_{\rm obs}=1$~yr of LISA operation. The wide band shows the result of the combined effect of EKL with two-body relaxation around SMBH binaries following N22, where the upper edge of each band assumes $\eta_{\rm bin}=1$ and $f_{\rm LISA}=f_{\rm EKL}=1$. The lower edge of the band instead adopts $\eta_{\rm bin}=0.5$, $f_{\rm LISA}=0.4$ and $f_{\rm EKL}=0.5$, the latter two consistent with N22 findings. The thin band in the Figure shows the estimated number of EMRIs from single SMBHs based on two-body relaxation, calculated from Eq.~(\ref{eq:number}). The  EMRI rate per single SMBH (i.e., $\gamma_{\rm relx}$) is estimated following \citet{Hopman+05}; see Eq.~15 in N22. The top panel of Figure \ref{fig:numbers} shows the boost in the EMRIs rate in the binary scenario relative to the single-SMBH case. 

As depicted in Figure \ref{fig:numbers}, even at the lower edge of the uncertain ranges shown, the combined EKL + two-body relaxation boosts the EMRI rate by more than an order of magnitude. Further, as expected, higher redshifts allow for a larger volume to be sampled and yield more EMRIs per unit $z$.

\section{LISA observability }\label{sec:LISA}

\subsection{Characteristic Gravitational Wave Strain}

Eccentric EMRIs emit GWs over a wide range of orbital frequency harmonics. The dimensionless GW strain $h(a,e,t)$ of an EMRI with a semimajor axis $a$ and eccentricity $e$ is the sum of the strains at each harmonic $f_n = nf_{\rm orb}$ where $f_{\rm orb} = \sqrt{GM_\bullet/a^3}$ \citep{Peters+63}. In other words the strain is:
\begin{equation}\label{eq:htimedomain}
    h(a,e,t)=\sum_{n=1}^{\infty} h_n(a,e,f_n)e^{i2\pi f_nt} \ ,
\end{equation}
where 
\begin{equation}
    h_n(a,e,f_n)= h_0(a)\frac{2}{n}\sqrt{g(n,e)} \ ,
\end{equation}
Here $h_0(a)$ is the instantaneous strain amplitude for a circular EMRI with SMBH mass $M_\bullet$ and stellar-mass BH mass $m$, at a luminosity distance $D_l$, whose value, averaged over sky location and GW polarization, is given by
\begin{equation}
           h_0(a) = \sqrt{\frac{32}{5}}\frac{G^2}{c^4}\frac{M_\bullet m}{D_l a} \ .
\end{equation}
Here $G$ is the gravitational constant, $c$ is the speed of light,  and $g(n,e)$ is defined as: 
\begin{eqnarray}\label{eq:gn}
g(n,e) &=&  \frac{n^4}{32}\Big[( J_{n-2} - 2 e J_{n - 1} + \frac{2}{n} J_n + 2 e J_{n+1} - J_{n+1})^2 \nonumber \\ 
&+& (1 - e^2)(J_{n-1} - 2 J_n + J_{n+2})^2 + \frac{4}{3 n^2}J_n^2\Big] \ ,
\end{eqnarray}   
where $J_i$ is the $i$th Bessel function evaluated at $ne$.  The characteristic strain 
 of a stationary EMRI $h_c(a,e,f)$, can be approximated using the Fourier transform of $h(a, e, t)$  measured for some time $T_{\rm obs}$ \citep{Kocsis+12}
 \begin{equation}\label{eq:hc}
     h_c(a,e,f) = 2 f \Tilde{h}(a,e,f) \sqrt{{\rm min}(T_{\rm obs},f_n/\dot{f}_n)}.
 \end{equation}
 The minimum term accounts for the (square root) of the number of cycles that can be observed at each frequency, determined by the smaller of the finite observation time and the time for the EMRI to chirp across the frequency band. 
 Lastly, $\tilde{h}(a,e,f)$ is the Fourier transform of Equation (\ref{eq:htimedomain}) over the observation time $T_{\rm obs}$
 \begin{equation}
\tilde{h}(a,e,f) = \int_0^{T_{\rm obs}} h(a,e,t) e^{-2 \pi i f t } dt \ , %[{\rm frequency}^{-1}]. 
\end{equation}
 and has units of Hz$^{-1}$.  
 Following \citet{OLeary+09}, \citet{Kocsis+12} and \citet{Gondan+18a}, we assume that there is negligible overlap between the harmonics, and thus the sum of squares of each element dominates the product of the sums in the Fourier transform. Note that a small fraction of systems ($<1\%$,{{ for a one-year observation, see below}}) is expected to evolve significantly in the LISA band and thus will require a more detailed analysis of their strain amplitude \citep[e.g.,][]{Barack+04}, which is beyond the scope of this work.  {{Note that evolving EMRIs tend to have a higher SNR; thus, the analysis below represents a conservative estimate for the resolvable sources and the stochastic background. }}

Figure \ref{fig:LISA} depicts the envelope of the characteristic strain of $2000$ EMRIs for $z\leq1$.  The value of $2000$ for the total number of systems is a result of the integral presented in the bottom panel of Figure \ref{fig:numbers}, see also Equation (\ref{eq:number}). We randomly draw $2000$ EMRI systems using the above mass function (see Appendix). Furthermore, we sample configurations in $a$ and $e$ from the fiducial $10^7$~M$_\odot$ simulations in N22\footnote{{{The N22 simulations were initialized with a cusp-like, \citet{Bahcall+76} distribution and a thermal distribution of eccentricities within the hierarchical limit. }}}, and rescale the results for different primary masses, keeping the periods similar\footnote{It was noted previously that EKL systems are scalable \citep[e.g.,][]{NaozSilk14,Naoz+19}. In these studies, the EKL timescale is kept constant. However, even when we relax the constant EKL timescale, the main contributor to the rate is the number of particles within the hierarchical limit \citep{Mockler+23,Melchor+23}. The number of BHs within the hierarchical limit is proportional to the mass of the primary via the $M-\sigma$ relation \citep[e.g.,][]{Tremaine+02}.  }. {{The EMRIs shown in Figure \ref{fig:LISA} are highly eccentric with an average eccentricity of $0.996$ and an average period of $\sim 5$ years. The realizations are chosen from the SMBH mass function described in the Appendix and yield an average SMBH mass of $1.6\times10^6$~M$_\odot$. We note that we kept the stellar mass BH fixed at $10$~M$_\odot$ throughout this {\it Letter}. These wide and eccentric systems result in a long GW merger timescale, with an average value of $2.3\times10^4$~yrs. Thus, only $1\%$ of the systems evolve within one year of observation time.}}

As clearly shown, many EMRIs lie below LISA's sensitivity curve \citep[the thick black line in the figure, calculated following][]{Robson+19}. The purple lines (above the LISA sensitivity curve) represent systems with signal-to-noise (SNR) larger than $8$, and they represent about $\sim 100$ ($\sim 40$) detectable sources, after $T_{\rm obs}=4$~yr ($T_{\rm obs}=1$~yr) from the first year of observations (recall that increasing $T_{\rm obs}$ for a stationary source increases it SNR)\footnote{{{ We note that an SNR of $8$ is lower by a factor of a few than some of the commonly adopted SNRs. However, these numbers are uncertain since LISA is yet to be launched.   }}}. After four years of observations, we expect a few hundred sources to become detectable.  

The SNR is defined \citep[e.g.,][]{Kocsis+12,Robson+19} as:
\begin{eqnarray}\label{eq:SNR2}
< {\rm SNR}^2(a,e)> & =& \frac{16}{5}\int \frac{ |\tilde{h}(a,e,f)|^2}{S_n(f)} df \nonumber \\
&=& \frac{16}{5}\int \frac{ |f\tilde{h}(a,e,f)|^2}{fS_n(f)} d\ln f.
\end{eqnarray}
We can thus  approximate the SNR for the eccentric source at the peak frequency as \citep[e.g.,][Xuan et al.~in prep.]{Kocsis+12}:
\begin{eqnarray}
   {\rm SNR} \sim \frac{ {\rm max}(h_c) }{\sqrt{ S_n\left(f_{\rm peak}\right)f_{\rm peak}}} \quad {\rm for}~P_{\rm orb} < T_{\rm obs} \ ,
\label{eq:SNR} 
\end{eqnarray}
where $S_n(f_{\rm peak})$ is LISA's sensitivity threshold.
In other words, among the expected $2000$ EMRIs from $z\sim1$ (see Figure \ref{fig:numbers}, bottom panel), we expect $2\%$ to be visible within LISA with $SNR>8$, per year, and the rest to contribute to confusion noise (depicted as thin grey lines).

Note that in some cases, we may resolve specific harmonics \citep[with a similar treatment described in][]{Barack+04Appr,Seto12,Katz+21,Xuan+23}. {{This detection of higher individual harmonics effectively can increase the SNR and potentially break the degeneracies between inferred parameters.}} 
Furthermore, the actual trajectory of the stellar mass BH, and its characteristic strain should be modified by Kerr geometry and special relativity in the case of spinning SMBH and wide systems, respectively, \citep[e.g.,][]{Yunes+08,Berry+13,Johannsen13,Schnittman15,Schnittman+18,Chua+17,Chua+21}. 
 
\subsection{Stochastic Background from Unresolved EMRIs}\label{sec:noise}

The population of EMRIs which do not individually rise to detectable levels constitute an unresolved GW background, which, in turn, can represent confusion noise for LISA.  We estimate the level of this confusion background ``noise'', 
$\sqrt{S(f) f}$ using two different methods.
The first of these is to generate a single Monte-Carlo realization of the sky with 2000 sources, using the results in the previous Section and following the approach of computing the corresponding GW background from a catalog of discrete sources by, e.g., \citet{Kocsis+11GW}, \citet{Nissanke+12} and \citet{Dvorkin+2017}. The second approach is to directly compute the expectation value of the local GW background using smooth functions describing the population of sources, following the approach by \citet{Phinney01}, generalized to eccentric binaries as in \citet{Enoki+07} and \citet{Toonen+09}.
%ZH4: I think the substantive division of methods 1 and 2 is that the first is for a discrete set of sources, which generate individual sky realizations, whereas the second is the "analytic" derivation of the expectation value of the background.  I divided the citations along this line, too, and added Irina Dvorkin's paper which also had Monte Carlo.
%SN4: sounds good

\begin{figure}
  \begin{center} %\vspace{-1.2cm} %\hspace{0.2cm} 
\includegraphics[width=\linewidth]{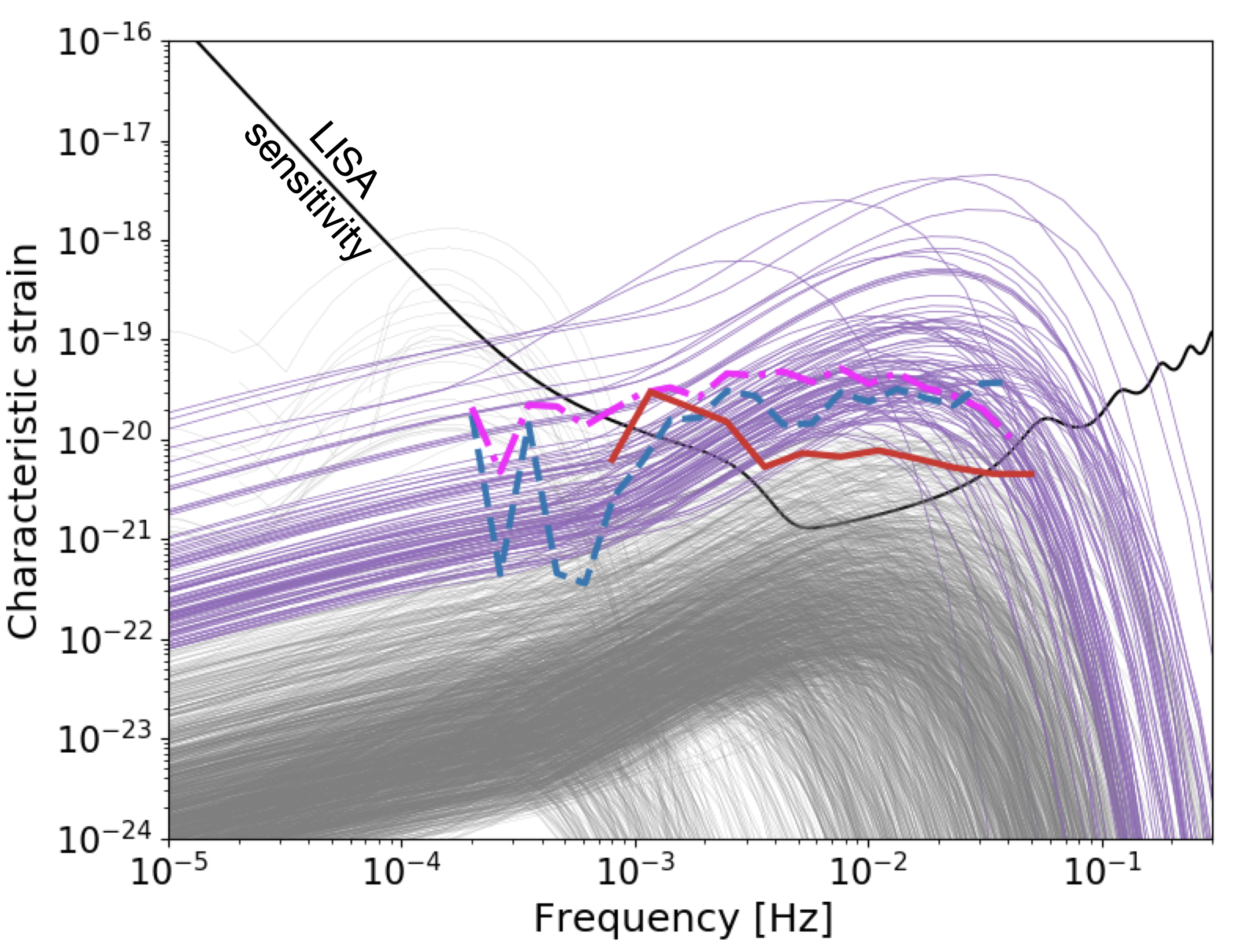}
  \end{center} %\vspace{-0.4cm} %\hspace{0.2cm}
  \caption{  \upshape {\bf{Characteristic GW strain and noise level for LISA}}. We show characteristic strains in the LISA band as a function of frequency. {{We show $2000$ randomly chosen EMRI systems and adopt $T_{\rm obs}=4$~yr.}} The thin {grey} lines show {{the unresolved systems, while the purple lines above the LISA sensitivity curve show the resolved systems with SNR$>8$}}. The thick red solid line shows $\sqrt{S(f)\times f}$, as described in the text. The dashed blue line shows $\sqrt{\sum {\rm max }(h_c)^2}$ for those $2000$ randomly chosen systems, where ${\rm max }(h_c)$ is the maximum characteristic strain of each system. The dashed dot magenta line shows $\sqrt{\sum (h_c)^2}$, including all the relevant frequencies of the strain for each realization.  The black line shows LISA's sensitivity curve following \citet{Robson+19}. 
  } \label{fig:LISA} %\vspace{-0.4cm}
\end{figure}

In the first method, we use the characteristic strain described above and estimate the noise as
\begin{equation}\label{eq:sumh}
\sqrt{S(f) f} \sim \sqrt{\sum_i {\rm max} [h_{c}(f)]^2} \ ,
\end{equation}
where ${\rm max} [h_{c}(f)]$ is the maximum characteristic strain (from Eq.~\ref{eq:hc}) for each EMRI. Equation (\ref{eq:sumh}) is motivated by the SNR definition, Eqs.~(\ref{eq:SNR2}) and (\ref{eq:SNR}). The summation is done in each observed frequency bin.  This result is depicted in Figure \ref{fig:LISA} as the thick (blue) dashed line.  For reference, the magenta, dot-dashed line curve in this figure shows the same calculation, except including the full frequency spectrum of each source rather than just the strain at the peak frequency. Both cases include only sources with SNR$<8$ and periods longer than one year.

In the second method, we follow \citet{Barack+04} where the one-sided spectral density $S(f)$ in the LISA band can be estimated as:
\begin{equation}\label{eq:Sf}
    S(f)= \frac{4 G}{\pi c^2 f^2}\frac{d\rho}{df} \ ,
\end{equation}
where $f$ is the observed frequency and $(d\rho/df)\Delta f$ is the total GW energy density observed today with frequencies of $(f,f+df)$, which can be estimated as
\begin{equation}
   \frac{d\rho}{df} \Delta f = \int_0^{z_{\rm max}} \Delta f_{\rm em} \frac{dt}{dz}\frac{\dot{\mathcal{E}}(f_{\rm em})}{1+z} dz = \int_0^{z_{\rm max}} \Delta f \frac{dt}{dz} \dot{\mathcal{E}}(f(1+z)) dz.
\end{equation}
Here $\dot{\mathcal{E}}(f_{\rm em})$ is the emitted spectrum of the GW sources (i.e., energy per unit comoving volume, unit proper time, and unit emitted frequency), and in the last step we used $f_{\rm em}=f(1+z)$.  The emitted spectrum can be approximated as
\begin{equation}
 \dot{\mathcal{E}}(f_{\rm em}) \sim \int \frac{d^2N}{dV d\ln M} \gamma(M) \Delta E_{\rm GW}(M,f_{\rm em}) d\ln M \ ,
\end{equation}
where $\Delta E_{\rm GW}(M,f_{\rm em})$ is the emitted GW energy as a function of mass in a frequency bin $(f_{\rm em},f_{\rm em}+df_{\rm em})$. We note that the $z>1$ events contribute negligibly to the total LISA signal. For each SMBH primary and a stellar-mass black hole of $m=10$~M$_\odot$, with a given semi-major axis $a$ and eccentricity $e$, the energy loss is \citep{Peters+63,Peters64}
\begin{equation}
    \frac{dE}{dt}= \frac{32 G^4 M_\bullet^2 m^2(M_\bullet+m)}{5 a^5 c^5} f(e) \ ,
\end{equation}
where $f(e)$ is the summation of all the harmonics, which gives: 
\begin{equation}
    f(e) =\left(1+ \frac{73}{24}e^2 + \frac{37}{96}e^4 \right)\left(1-e^2\right)^{-7/2} \ .
\end{equation}
We use the fiducial system of a $10^7$-$10^9$~M$_\odot$ SMBH binary in N22. We re-scale the EMRI's semi-major axis for each mass based on this fiducial system. For simplicity, we assume that most of the energy is emitted at the peak frequency  $f_{\rm peak} = f_{\rm orb} \sqrt{(1+e)/(1-e)^3}$, where $f_{\rm orb} = \sqrt{GM_\bullet/a^3}$. Note that each $dE(a,e)/dt$ has its characteristic $f_{\rm peak}$.

In order to find $\Delta E_{\rm GW}(M,f_{\rm em})$, we draw $4000$ random EMRI configurations (i.e., a set of $a$ and $e$) from the fiducial simulation and rescale for each $M_\bullet$. 
Then, we find the fraction of the cases in which a $dE/dt$ value appeared in the frequency bin $f_{\rm em}=f_{\rm peak}(a,e)$. This process, thus, roughly estimates the energy loss probability of an EMRI for a given primary mass in a frequency bin. We then estimate $\Delta E_{\rm GW}(M,f_{\rm em})$ for a typical EMRI, over its lifetime, at mass $M_\bullet$, as  
\begin{equation}
   \Delta E_{\rm GW}(M,f_{\rm em}) \sim  \frac{1}{N_{\rm real}}\sum_{\rm real} \left(\frac{df_{\rm em, peak}}{dt}\right)^{-1}\frac{dE(a,e,f_{\rm em,peak})}{dt}  \ ,
\end{equation} 
where the summation represents averaging over the (a,e) distribution at fixed mass and frequency, and we assume each EMRI, drawn from the N22 distribution in (a,e), subsequently evolves purely due to GW emission, i.e., $df_{\rm em, peak}/{dt}$ is given by (shown here for completion):
\begin{equation}
\frac{df_{\rm em, peak}}{dt} = \frac{\partial{f}_{\rm em, peak}}{\partial{a}}\frac{da}{dt} + \frac{\partial{f}_{\rm em, peak}}{\partial{e}}\frac{de}{dt} \ ,
\end{equation}
where $f_{\rm em, peak}$ is defined above and $da/dt$ and $de/dt$ are taken from \citet{Peters+63} and \citet{Peters64}.

Figure \ref{fig:LISA} depicts the confusion noise estimated following the above prescriptions. Specifically, the thick red line shows $\sqrt{S(f)f}$. This figure suggests that in an idealized case, in which all galaxies host one SMBH binary, the noise level due to unresolved EMRIs covers most of the LISA band between (3-30)~mHz at the few$\times 10^{-20}$ level.  We note the consistency between the two methods used here to calculate the noise level. The first method (i.e., $\sqrt{\sum {\rm max }(h_c)^2}$) is cruder but appears to be consistent with the analytical calculation (thick red line). 

We suggest that the spectrum of this stochastic EMRI background noise may be ``jagged'' due to the finite number of EMRIs within the observable volume. As shown in Fig.~\ref{fig:numbers} $\mathcal{O}(1000)$ EMRI systems may exist in nature within $z\sim 1$, each contributing primarily to different frequency bins. Thus, while in principle, we could sample from a larger simulated sample and create a smoother spectrum, this may be misleading. The precise shape of the jagged noise spectrum is uncertain, as it varies from one set of realizations to another.

\section{Discussion and Astrophysical Implications}\label{sec:diss}

The hierarchical nature of galaxy formation implies that SMBH binaries or even higher multiples should be the common outcome of galaxy mergers (see Figure \ref{fig:cartoon} for an illustration). Furthermore, since galaxies' minor mergers are common, a large SMBH mass ratio in SMBH binaries may be a natural consequence \citep[e.g.,][]{Ricarte+16,Rashkov+14}. 
%\citep[e.g.,][]{Rodriguez-Gomez+15,Pillepich+17,Paudel+18,Micic+23}
It was recently shown that these systems can yield a high EMRI rate per galaxy \citep{Naoz+22}. 
Here we demonstrated that this high rate implies that thousands of SMBH-stellar mass BH GW sources may exist within the local Universe (redshift of unity); see Fig.~\ref{fig:numbers}. Among these EMRIs, a few hundred may be detectable with SNR$>8$ by LISA during its lifetime, and the rest, unresolved ones, accumulate to a high stochastic background noise level. As depicted in Figure~\ref{fig:LISA}, the level of this background can be an order of magnitude above LISA's sensitivity limit.

We note that the stochastic background from EMRIs described above is analogous to the background from Galactic dwarf binaries. This white dwarf background has been estimated (assuming circular binary inspirals) and is known to reduce LISA's sensitivity at $\sim 2$~mHz \citep[e.g.,][]{Nissanke+12,Robson+19}. The reduced sensitivity manifests as a ``bump'' in the LISA sensitivity curve, increasing it by a factor of a few.   Here we find that the EMRI background may result in a similar, but potentially even higher ``bump'' at $\sim 3-30$~mHz. However, unlike the nearly ten billion Galactic white dwarfs that produce a smooth feature, here, we find that the EMRI background may result in a noisy, jagged curve.
%
%This jaggedness is a result of the combination of the discrete nature of EMRIs (of which there are much fewer than white dwarfs), as well as
This jaggedness is a result of the combination of the discrete nature of EMRIs (of which there are much fewer than white dwarfs), as well as their nonzero eccentricities, which spreads them over a wider range of frequencies compared to circular inspirals, causing even fewer sources to contribute per frequency bin.

{{The background noise presented here is much higher than previous, single-SMBH estimates. Specifically, previous analyses of the EMRI background found that its expected level is very close, or slightly above LISA's sensitivity curve, which includes the unresolved Galactic WD binaries \citep[][]{Pozzoli+23,Barack+04}. \citet{Bonetti+20}'s optimistic model is slightly below the conservative background estimates presented here. Further, their optimistic background noise estimate drops below the LISA sensitivity curve at around $0.01$~Hz, while the unresolved background presented here is roughly constant.}}

\begin{figure}
  \begin{center} %\vspace{-1.2cm} %\hspace{0.2cm} 
\includegraphics[width=\linewidth]{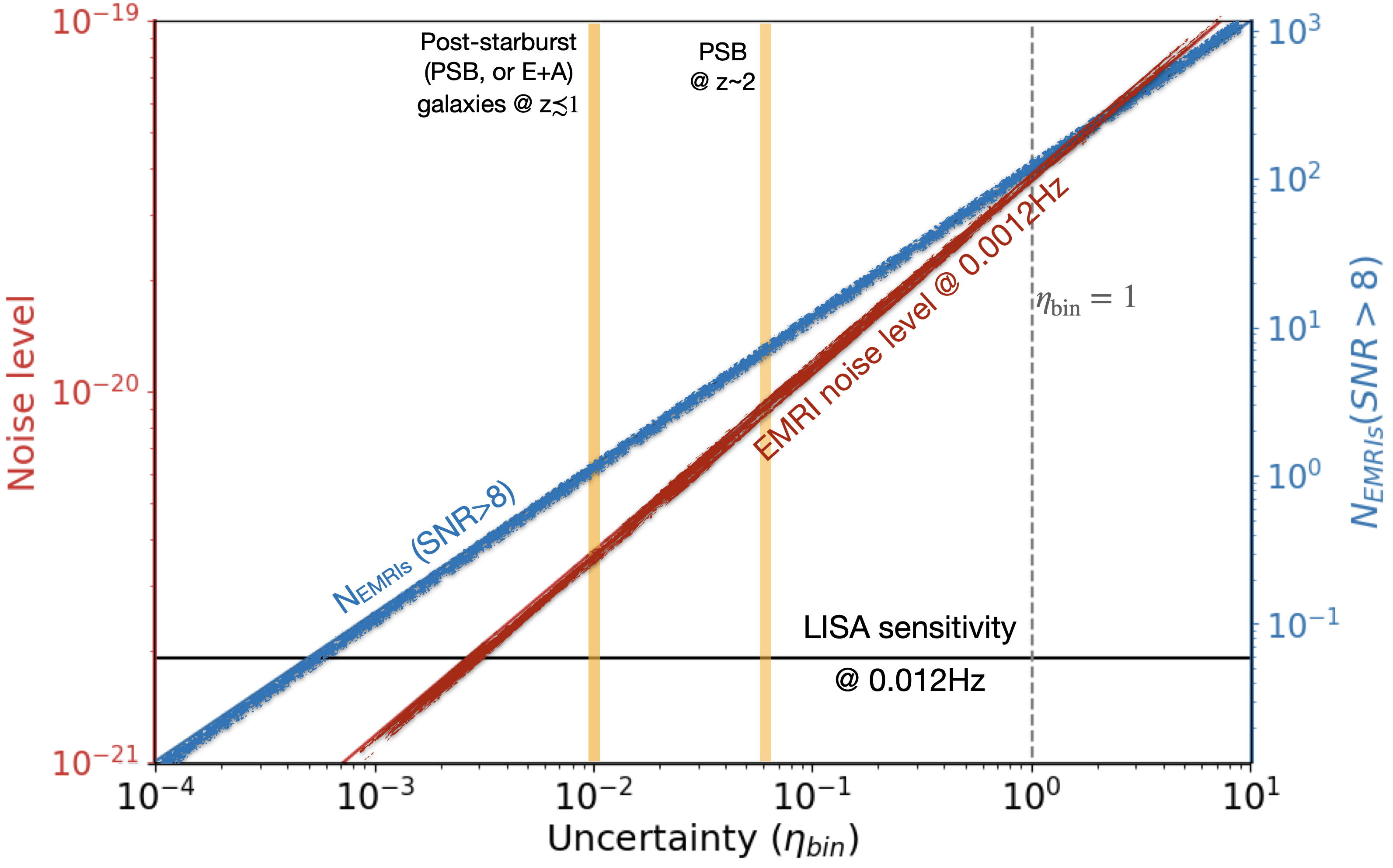}
  \end{center} %\vspace{-0.4cm} %\hspace{0.2cm}
  \caption{  \upshape {\bf{The astrophysical consequences of binary EMRIs}}.
    The x-axis shows the SMBH binary fraction, $\eta_{\rm bin}$, which incorporates all of the uncertainties mentioned in this work. This term can be larger than unity if SMBHs in galactic nuclei typically have multiple companion SMBHs/IMBHs. The left y-axis (red) shows the EMRI noise level at $\sim 0.012$~Hz estimated from the noise calculation described in Section \ref{sec:noise} and shown in Figure \ref{fig:LISA}, as a function of $\eta_{\rm bin}$. LISA's sensitivity at $\sim 0.012$~Hz (black, horizontal line) is overplotted. This frequency corresponds to the largest difference between the EMRI noise level and the LISA sensitivity curve. The right y-axis (blue) shows the estimated number of EMRIs with SNR$>8$, and $T_{\rm obs}=4$~yr, as a function of $\eta_{\rm bin}$.  The thick vertical yellow lines mark the approximate fractions of post-starburst (PSB) galaxies in the local Universe ($z\lesssim 1$) and at $z\sim 2$.   } \label{fig:CartoonPlot}
  \vspace{0.5\baselineskip}
\end{figure}

We capture the uncertainties in the above estimation with the parameter $\eta_{\rm bin}$. This parameter describes the average number of light companions around a central SMBH with overlapping spheres of influence,
thus, in principle, $\eta_{\rm bin}$ can be larger than unity. Additionally, $\eta_{\rm bin}$ may be degenerate with other uncertainties, including (but not limited to) the fraction of stellar-mass BHs in nuclear star clusters, the fraction of plunges vs. slowly inspiraling EMRIs (denoted as $f_{\rm LISA}$, above), the fraction of stellar-mass BHs that will be driven toward the primary in the aforementioned channel  (denoted as $f_{\rm EKL}$, above), and the hardening time of SMBH binaries \citep[although an SMBH undergoing a merger was also suggested to yield a high EMRI rate, e.g.,][]{Mazzolari+22}. However, unlike the SMBH binary fraction, $f_{\rm LISA}$, and $f_{\rm EKL}$  may be constrained using the N22 simulations. In particular, the latter is an estimate of the fraction of particles that crossed the hierarchical limit, and the former is derived from estimating the fraction with periods shorter than $\sim 10$~yrs; see Appendix~B in N22. Notably, even at the low end of their expected range, these parameters still significantly enhance the EMRI rate (by over an order of magnitude, see the top panel in Fig.~\ref{fig:numbers}).

We suggest that LISA observations can be used to constrain the astrophysical nature of SMBH binaries —- particularly the average number of companions of SMBHs. This is illustrated in Figure \ref{fig:CartoonPlot}, where the x-axis is $\eta_{\rm bin}$. The left y-axis (red) depicts the noise level from the unresolved EMRIs at $\sim 0.012$~Hz, shown as the thick red line (labeled). This frequency corresponds to the maximum noise compared to the LISA's sensitivity threshold (depicted in the Figure as a black horizontal line at $\sim 0.012$~Hz). Equation (\ref{eq:Sf}) combined with Equation (\ref{eq:gamma}) suggests that the noise level $\sqrt{S(f) \times f} \sim \sqrt{\eta_{\rm bin}}$. Thus, measuring this stochastic background noise level by LISA can be used to constrain $\eta_{\rm bin}$. Similarly, the number of LISA-detected EMRIs can be used to constrain the incidence rate of SMBH binaries. This is illustrated by the right y-axis of Figure \ref{fig:CartoonPlot}, where we show the number of systems with SNR$>8$ detectable by LISA per year. This is estimated from the calculations presented in Figure \ref{fig:LISA}, where $\sim 6\%$ of the systems from $z\lesssim 1$ have SNR$>8$.  As illustrated in Figure \ref{fig:CartoonPlot}, the combination of the number of individually detected EMRIs and the level of the unresolved background should yield strong constraints on the SMBH binary population.

{{We highlight that the EMRI rate predicted from SMBH binaries is much higher than the EMRI rate based on single SMBHs  (see Figure \ref{fig:numbers}). Thus, for non-negligible values of $\eta_{\rm bin}$, the single-SMBH EMRI population has a minor contribution to the background noise. Disentangling between the binary and single channels for a resolved EMRI may be possible if the EMRI can be localized and the mass of the EMRI-SMBH is found to deviate (i.e., falls well below) from the observed or inferred central SMBH mass. Such a possibility may indicate that the stellar-mass BH merged with a smaller, hidden SMBH. Another possibility to differentiate the single and binary channels may be through their respective electromagnetic signatures. For example, a concurrent detection of an EMRI and a TDE \citep[e.g.,][]{Melchor+23} may help constrain the mass of the disrupting SMBH, adding an independent measure of the mass.  }}

Notably, taken at face value, if every SMBH at the center of a galaxy hosts at least one smaller SMBH companion, the same channel that produces a high EMRI rate will result in a high TDE rate. Perhaps even higher than the observed TDE rate per galaxy \citep[estimated as $10^{-5}-10^{-4}$, per year, per galaxy, e.g.,][]{French+20}.
It has been shown recently that the combined EKL + two body relaxation indeed results in a high TDE rate \citep{Melchor+23}. Furthermore, it can be used to detect the lighter SMBH \citep{Mockler+23}. Interestingly, the TDE rate predicted from the combined EKL + two body relaxation channel is consistent with the TDE rate observed in post-starburst (PSB) galaxies \citep{Melchor+23}. Remarkably, the observed TDE rate in post-starburst galaxies is higher than the averaged observed TDE rate \citep[e.g.,][]{vanVelzen+20}. 
 Thus, these results suggest that SMBH binaries near the sphere of influence of each other may be a typical outcome of PSB galaxies. This inference appears to be consistent with the recent NANOGrav result, suggesting that high-mass SMBH binary mergers take place within the first $1$~Gyr of galaxy mergers \citep[e.g.][]{Antoniadis+23NaNoGravSMBHBin}. 

 Thus, in Figure \ref{fig:CartoonPlot}, we overplot the estimated PSB galaxy fraction at $z\lsim 1$ and $z\sim 2$ \citep{Bergvall+16,Wild+16,Belli+19,DEugenio+20,French21}.  Thus, if SMBH binaries exist only in PSB galaxies\footnote{In this case, the rest of the population may have quickly merged or stayed in the galaxy's bulge. Alternatively, the light SMBH did not hold to its nucleus as it sank in.}, then the PSB galaxy fraction may indicate the fraction of SMBHs with companions, i.e., $\eta_{\rm bin}$. The consistency of the TDE rate with the EKL + two body relaxation \citep{Melchor+23} estimate may hint at such a scenario.    As shown by this figure, limiting the stochastic background noise from binaries to originate only from PSB galaxies will still create a significant level of noise confusion in LISA in its most sensitive frequency band. In other words, a noise level $2-8\times 10^{-20}$ (for $z\sim 1-2$) may imply that the EMRIs' origin is consistent with PSB galaxies.  

\section{Conclusions}\label{sec:con}

In this paper, we demonstrated that the high EMRI rate predicted in SMBH binaries \citep{Naoz+22} can result in thousands of GW sources. A considerable fraction of these sources may be detected individually by LISA (estimated as a few hundred sources for a four-year LISA mission lifetime). The accumulated, unresolved sources, in turn, may accumulate to a significant unresolved stochastic background, reaching an amplitude that is an order of magnitude above LISA's sensitivity threshold between $\sim(3-30)$mHz. Finally, we suggest that in a simple interpretation, LISA observations of individually detectable EMRIs and the EMRI noise levels can be used in combination to constrain the prevalence of SMBH binaries in galactic nuclei.

\acknowledgments
We thank Zeyua Xuan, Brenna Mockler, and Suvi Gezari for useful discussions. We also thank the anonymous referee for their useful comments. 
S.N. acknowledges the partial support from NASA ATP 80NSSC20K0505 and from NSF-AST 2206428 grant as well as thanks Howard and Astrid Preston for their generous support.

\appendix
\section{The Mass function}\label{sec:massfunc}

To SMBH mass function is adopted from \citet{Merloni+08}. We develop a simple fitting formula for their mass function as a function of redshift and BH mass. We find that the SMBH mass function can be estimated by:
\begin{equation}
    \log_{10} \left( \frac{dN}{dV d\log M}(z)\right) = a_1(z) (\log_{10} M)^4 +  a_2(z) (\log_{10} M)^3 +  a_3(z) (\log_{10} M)^2 +  a_4(z) \log_{10} M + a_5(z) \ ,
\end{equation}
% p1(z)*x^4 + p2(z)*x^3 + p3(z)*x^2 + p4(z)*x + p5(z) and p_i(z) 
where $a_i(z)$ is fitted to be:
\begin{equation}
    a_i(z) = p_{i1}z^4 + p_{i2} z^3 + p_{i3} z^2 + p_{i4} z + p_{i5} \ ,
\end{equation}
and
\begin{equation}
p_{ij} = \left(
\begin{array}{ccccc}
 0.023365 &  -0.1031 &   0.158098 & -0.080575 & -0.009329 \\
-0.747931 &   3.335781 &  -5.154087 &   2.683332 &  0.176503 \\
 8.8832& -40.0616&  62.3869&-33.2041&  -0.9996\\
-46.4664 & 211.9867 & 333.0359 &  181.6464&   0.622461\\
 90.3990 & -417.6483 &  663.5951 & -373.2478&   4.26139
\end{array}\right) \ .
\end{equation}

\begin{figure}
  \begin{center} %\vspace{-1.2cm} %\hspace{0.2cm} 
\includegraphics[width=0.5\linewidth]{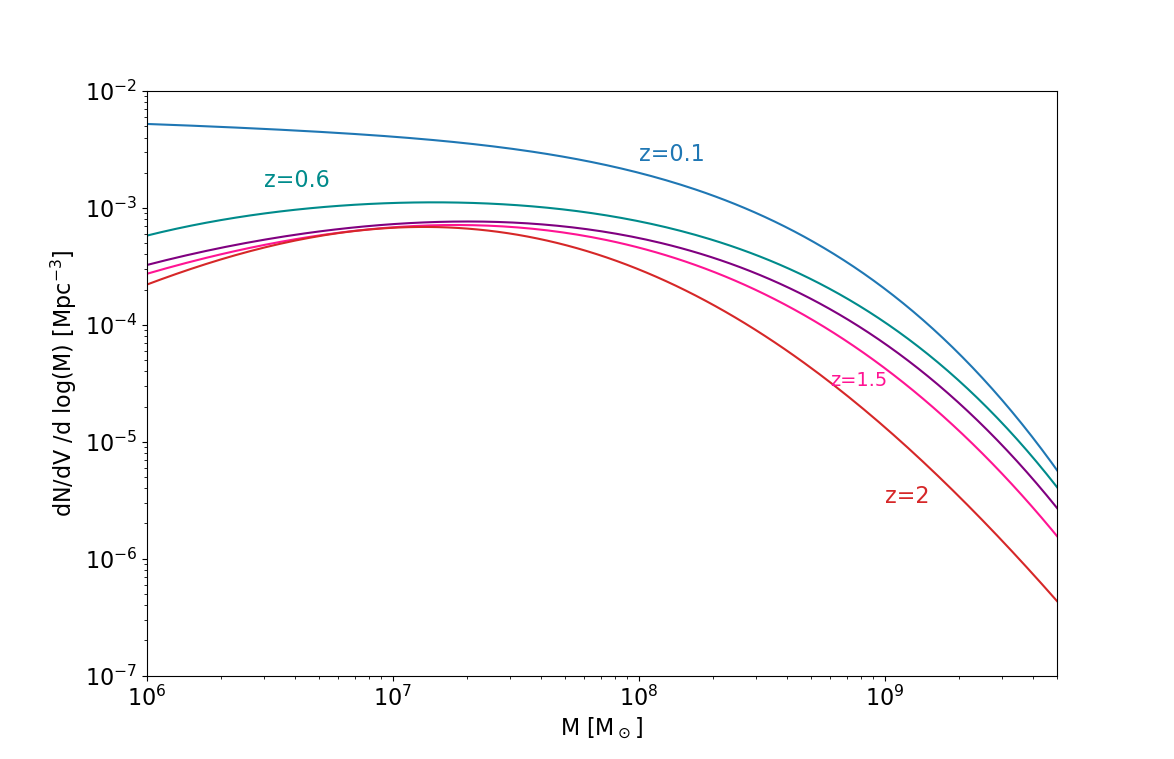}
  \end{center} %\vspace{-0.4cm} %\hspace{0.2cm}
  \caption{  \upshape {\bf{Examples of the mass function fits used in this calculations}}. These are fits for \citet{Merloni+08}; see Appendix for details.  } \label{fig:massfunction} %\vspace{-0.4cm}
\end{figure}

The results from this fit are depicted in Figure \ref{fig:massfunction}.

\vspace{\baselineskip}

\bibliographystyle{hapj}
%\bibliography{papers2}
\bibliography{Binary}
\end{document}